# Citizen Perspectives on Necessary Safeguards to the Use of AI by Law Enforcement Agencies

Yasmine Ezzeddine[1]
 *CENTRIC*
*Sheffield Hallam University*
*Sheffield, UK*
y.ezzedine@shu.ac.uk

Petra Saskia Bayerl
*CENTRIC*
*Sheffield Hallam University*
*Sheffield, UK*
p.s.bayerl@shu.ac.uk

Helen Gibson
*CENTRIC*
*Sheffield Hallam University*
*Sheffield, UK*
h.gibson@shu.ac.uk

**ABSTRACT.** In the light of modern technological advances, Artificial Intelligence (AI) is relied upon to enhance performance, increase efficiency, and maximize gains. For Law Enforcement Agencies (LEAs), it can prove valuable in optimizing evidence analysis and establishing proactive prevention measures. Nevertheless, citizens raise legitimate concerns around privacy invasions, biases, inequalities, and inaccurate decisions. This study explores the views of 111 citizens towards AI use by police through interviews, and integrates societal concerns along with propositions of safeguards from negative effects of AI use by LEAs in the context of cybercrime and terrorism.

*Keywords:* Artificial Intelligence; Law Enforcement Agencies; Safeguards; Citizen Perspectives; Police

## 1. Introduction

The role of Artificial Intelligence extends beyond improving the security and safety of citizens, particularly against cybercrime and terrorism, to anticipate and recognize criminals' increasing employment of AI tools (Trend Micro Research, 2020). In fact, societies are embracing new forms of reality amplified by machine learning and use of AI (Mann, 2017), where every detail of daily routines is captured, stored, and digitalized. And once this information is distributed in the system, recalling it is nearly impossible (Petersen & Taylor, 2012). Hence, advancing the measures taken for public protection is imperative for enhancing general safety and security (Macnish, 2021). Doubtless, algorithms and data analytics are playing an increasing role in all aspects of society including the policing and security services (Babuta & Oswald, 2020) to the extent that policing through social media has been explored by several Law Enforcement Agencies (LEAs) worldwide. However, the expansion of data collection efforts and AI use continues to trigger uncertainty around its ethical and moral implications (Lyon, 2002), especially with respect to recent technologies involving the web, video monitoring and algorithmic decision-making warrants the need for critical evaluation of the inevitable psychological consequences (Stoycheff, et al. 2020) contributing to the skepticism around AI use by police.

---

[1] **Corresponding Author:** Yasmine Ezzeddine – S1 1WB – y.ezzeddine@shu.ac.uk

Additionally, the conflict mounts between the facilitated admittance that these technologies offer, and the "fear of contact" emanating from alliances with independent bodies of the private sector (Trottier, 2017, p. 475), coupled with the lack of evidence around efficiency of algorithmic-based decisions, accuracy, fairness, and risks of predictive policing leading to discrimination and inequality (Bushway, 2020; Quattrocolo, 2020; Završnik, 2020).

Similarly, amongst the numerous challenges facing AI implementations for LEAs as well the private sector is to determine how to capitalize on AI capabilities in response to changing safety and security challenges while ensuring responsible use. In fact, the AP4AI[2] Framework incorporating 12 Accountability Principles of AI laid the foundation for a "healthy balance between the need to innovate practices and enhance capabilities (…) on one hand, and the legitimate expectations by citizens that police work is conducted lawfully proportionality and in pursuit of a legitimate aim" (Akhgar et al., 2022, p.5).

Nevertheless, the scarce and limited systematic knowledge around citizen perceptions of safeguards inspired us to complement the existing insights around resistance to LEAs' data collection and use of AI, while satisfying the theoretical gaps around different types of safeguards are often considered in ethical and legal perspectives but not from the societal perspective of citizens.

Hence, the central theme of this research comprises an investigation of citizen propositions of necessary safeguards that can protect them from the potential negative effects incurred in AI use by police. In other words, this study focuses on engaging with citizens as not only beneficiaries of such innovations, but also key players in legitimizing the deployment of AI tools, since in the absence of citizens' approval and support, AI implementation can face the negative implications of chilling effects (Stoycheff, 2016), fear of contact (Marx, 2009; Marthew & Tucker, 2017), countertractions against police (Bayerl et al., 2021) and even national and international movements opposing its deployment (Montag, et al. 2021; Reclaim-your-Face[3] campaign). This warrants the investigation into citizens' perspectives to AI implementation as well as their propositions of safeguards to the potential negative consequences of incorporating AI technologies into LEA security practices.

---

[2] **AP4AI:** Accountability Principles for Artificial Intelligence: www.ap4ai.eu
[3] **Reclaim Your Face:** Ban Biometric Mass Surveillance! (n.d.). Reclaim Your Face. https://reclaimyourface.eu/

## 2. Methodology

This study adopts a qualitative approach aiming to better understand and integrate citizens' perspectives about data collection and AI use by LEAs. Therefore, semi-structured interviews were conducted to provide in-depth insights and elucidations into necessary safeguards, allowing citizens to elaborate on their subjective views and experiences.

### 2.1. Sample

As part of EU funded project AIDA we have conducted in eight countries. No specific criteria in terms of demographics were sought, except for at least 16 participants from the 'general population' of each participating country (above 18 years). The rationale behind this open choice was driven by two considerations: pragmatism, facilitating access to citizens, and guided interest, allowing partners to choose groups that are of interest to them. In total, 111 interviews were conducted with 44 males and 69 females were interviewed, the youngest participant being 19 years old and the oldest 83 years old.

The sample was varied in terms of experience with cybercrime and terrorism. About 58.2% indicated that they had no personal experience with either, while 34.5% reported experience with cybercrime or other incidents online (e.g., phishing, identity fraud, hacked email account). Only 4.5% had experience with terrorism (e.g., car attack in hometown). This reflects the considerable heterogeneity of experience and safety perceptions and therefore does not seem to be biased in a specific way towards citizens with high/low experience or specific attitudes towards safety.

### 2.2. Data Collection

Participants were recruited through researchers handling interviews in each of the countries. Semi-structured scenario-based interviews were conducted, in the respective country language, along pre-defined themes categorized in three main topics: "understanding of AI and acceptance conditions", "perception of safety with respect to terrorism/cybercrime and societal resilience", and "citizen reactions". This paper reports the findings from topic 3 related to safeguards, obtained particularly from responses to the question of *"What should police forces do to safeguard you from negative effects of AI systems?"* Interviews were either conducted online or face-to-face and audio recorded. For all interviews, Subsequently, signed consent forms, interview recordings, and English summaries or verbatim transcript in the country's language were provided. The latter were translated to English using an online software followed by proof-reading.

### 2.3. Data Analysis

Our analytic approach followed thematic and content analytic principles (Auerbach & Silverstein, 2003; Krippendorff, 2003) to identify the main topics and themes in the collected data starting with open or initial coding (Charmaz, 2006). Initial codes were then clustered into high-order categories per main topic. The categories identified for safeguarding measures are presented in the findings.

### 2.4. Ethics

Several steps were taken to ensure data collection adhered to relevant ethics requirements. Firstly, the study received approval by the ethics committee of Sheffield Hallam University. Secondly, the interviews started by presenting participants with an information sheet to clarify the context of the project, details of data handling, participants' rights, and legal basis for the study. Thirdly, participants were only asked for basic personal information (gender and age). They were further reminded of their right to not answer demographic questions if they did not feel comfortable to do so, which several participants rightfully used. All data was anonymized before analysis and interpretation.

## 3. Findings

Overall, participants produced 113 recommendations for safeguards they deem necessary to protect from potential negative effects of AI. In-depth evaluations of these revealed complex attitudes that can be clustered into the following areas (see Fig.1):

1) Educational safeguards: for citizens and LEAs
2) Technical safeguards: Regular Evaluations, anonymization of collected data
3) Legal safeguards: National and international regulations and independent agencies
4) Human safeguards: selectivity in employing AI-handling staff and importance of human validation of AI findings and decisions
5) Privacy safeguards: Limited data collection and requests for consent
6) Stop use of AI: use traditional means
7) Inevitable vs. no negative effects: either not foreseeing any negative effects or assuming no safeguards can protect from them

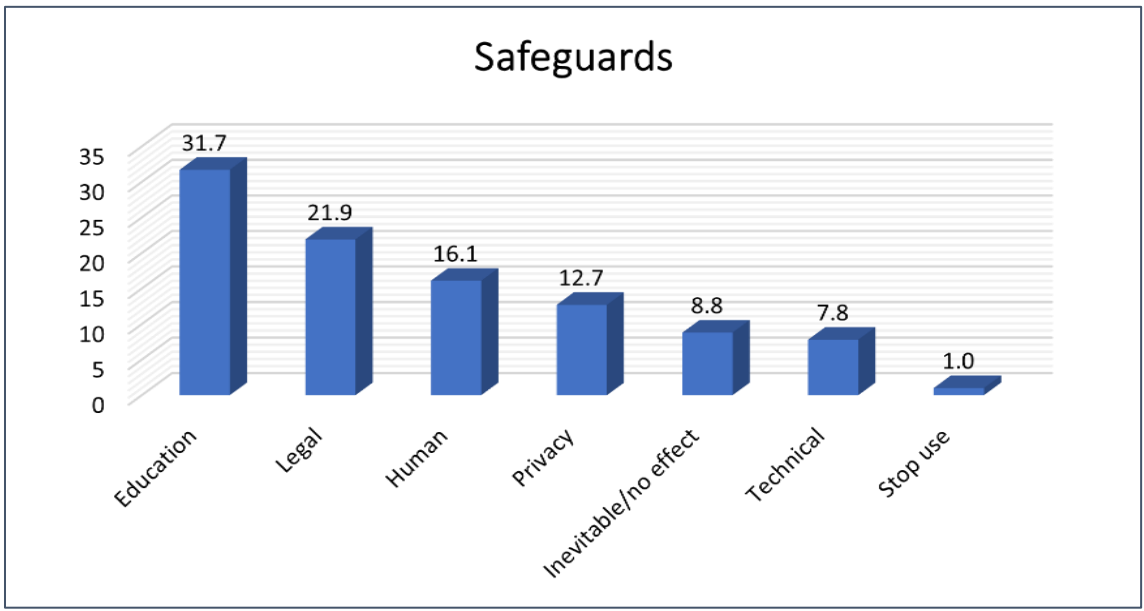

*Figure 1: Different types of safeguards suggested by interviewees (percentage of statements)*

### 3.1. Educational safeguards and transparency

To safeguard citizens from potential negative effects of AI, participants emphasized the importance of **education of LEA staff as well as citizens**. LEA personnel should receive enhanced and ethical training on how to handle AI tools. In extension, some participants proposed that tools should only be used by experienced LEA members while stressing the need for collaborations with outside experts, e.g., "*law enforcement are not scientists, and so they might make a mistake and that's not good, so I wish they have like a team that they've got people who are helping them*" (UK-02). Additionally, participants emphasized civic education, particularly educating children in schools around the process of data collection, the actual purposes of AI use, and the percentage of decision-making to which AI contributes. This links to another important aspect participants referred to repeatedly, namely **transparency**. This should include the sharing of positive outcomes of AI use and developing campaigns that showcase positive scenarios of AI use in criminal investigations or using an open-source platform to increase community trust.

### 3.2. Technical and AI specific safeguards

Technical and AI-specific safeguards were a second recurring theme., Participants referenced **regular assessments and evaluation** of AI tools (including impacts on crime rates), and the importance of controlling biases and detecting errors in algorithms to prevent their re-occurrence.

Additionally, participants raised the importance of adopting **technical safeguards** to ensure that the information is properly "anonymized", and not being used for other purposes or by other companies. Moreover, interviewees noted that criminals and terrorists also use AI tools which is why it is important that LEAs "*always trying to keep ahead*" (UK-11) of criminals by ensuring their AI tools are as up to date as possible.

### 3.3. Legal Safeguards: Frameworks and Policies

Participants further asked for legal frameworks, **national and international regulations** and policies that ensure "*well-defined and well-enforced limits for what LEAs are allowed to do with the data*" (NL-06). This may require preventive and punitive regulations to the misuse of AI, whether by LEAs or the private sector. Participants also stressed the need for communication between different law enforcement agencies and cooperation between different countries. In other words, AI should be monitored and supervised, preferably by an **independent agency** or unbiased third-party or government that regulates AI, enforces legislation, monitors data collected and ensures data is stored safely and within legal timeframes.

### 3.4. Human Safeguards: Avoiding errors and Biases

Participants further suggested **selectivity** and proper **vetting** of staff involved in roles touching on AI data collection/handling while ensuring ethnic diversity and gender equality to minimize biases. Additionally, the importance of the **human component in decision making** was perceived as crucial in monitoring and verifying AI decisions. Participants repeatedly stressed the importance of final decisions and interpretations to be done by humans since machines "cannot replicate humans". Another singular suggestion to reduce the possibility of biases was: "*use an AI system that is developed without unfair bias in accordance with the applicable laws after much research on its development and with the involvement of citizens during the research*" (GR-14).

### 3.5. Privacy Safeguards: Regulated Data Collection

With respects to automatic identification and random monitoring of members of society, participants expected LEAs to not invade citizens' privacy and freedoms for the sake of security. Instead, there should be **limitations on data collection**, such as "*whether there are sensitive areas of society where perhaps it shouldn't be used, particularly around our homes*" (NL-10). Participants strongly believed that LEAs should not waste time and resources collecting information from 'everyone'; instead, they should use alternative means to obtain data. Hence, they stressed that LEAs should not collect more information than they actually need, calling for regulation of data collection and judicial control of obtained information (see 4.3). Alternatively, LEAs should ask for individuals' **consent** prior to data collection, or hide/blur faces of people not suspected in a certain footage, as well as dispose of non-relevant information as soon as possible.

### 3.6. Stop use of AI

The most extreme position was expressed by a small group of statements that expressed total opposition to the implementation of AI. To them, the best safeguard was "*maybe not using Artificial Intelligence*" (PT-06). Instead, for this group LEAs are expected to work harder in the traditional way.

### 3.7. Inevitable vs. no negative effects to require safeguards

On the contrary, a small number of statements suggest that some participants did not perceive any negative effects to be safeguarded against. Few even encouraged LEAs to expand AI use. Other viewpoints indicated that there will always be negative effects which cannot be eliminated, either because LEAs lack the expertise on how to safeguard against negative consequences effectively, or due to the impossibility to control social media platforms collecting data. Hence, safeguards would either not be needed or not possible.

## 4. Discussion

With most eyes set on AI implementation in almost all aspect of modern life, calls for frameworks, regulations and safeguards are equally arising. The UN Rights Chief stressed the importance for safeguards in face of the "undeniable and steadily growing impacts of AI technologies" and the need to "protect and reinforce all human rights in the development, use and governance of AI as a central objective" (Geris & Wellington, 2021, n.p.). This coincides with the European Parliament's Press release that stressed the importance of subjecting AI use in policing to "strong safeguards and human oversight" (European Parliament, 2021). All of this resonates with participants' statements around educational, technical, legal, and human safeguards. It appears that citizens are not entirely against AI use, in fact, they are only discreet around its implementation, and have heightened, yet justifiable, concerns around the impact of AI on their privacy and overall quality of life.

One of the unique aspects of this study is the intersection it provides between citizens' attitudes towards AI, safety perceptions and propositions of safeguards, particularly as suggested by non-AI experts. This reveals that common citizens possess a basis of knowledge and understanding of AI and the inevitable consequences incorporated in its design. However, that did not create rejection of the entire tool. In fact, citizens were supportive of AI implementation as a tool that can safeguard societies, especially in the current modern era, if it adheres to strict rules and is monitored by trained and trustworthy individuals. This was reflected in the overall number of propositions on the need for educational and legal safeguards in the implementation of AI tools by LEAs, compared to the surprisingly lower rates on privacy safeguards and requests to stop using AI altogether.

Another beneficial aspect of these propositions lies in their potential to adapt existing AI tools and shape prospective designs to account for citizen perspectives, which may in turn reduce resistance and counterstrategies to data collection and AI implementation and hence, enhance the feasibility of information gathering while safeguarding the quality of collected data.

Other ramifications of this also involve financial implications by safeguarding data collection as a massive source of income (Deulkar & Gupta, 2018), all of which can benefit from potential application of safeguard propositions put forth by citizens participating in this study.

### 4.1. Limitations and Future Work

The heterogeneity of participants across countries allowed us to obtain a highly diverse set of experiences and perspectives. Yet, the small number of interviews per country inhibits an analysis of subgroups. However, this qualitative approach can outline the personal, individual perspectives of citizens across contexts which in turn, reflects the richness of citizen views, while displaying similarities in opinions and expectations alongside personal motivations and reasoning. Future research can build on these findings to include the demographic groups most and least critical towards LEAs' use of AI in any country.

### 5. Conclusion

This study provides an in-depth evaluation of citizens propositions of safeguards to LEAs' use of AI. In brief, AI use should be justified, legitimate and only used for declared purposes. Other safeguards included the avoidance of biases through appropriate AI design, supervision and legal framework, regular evaluations, transparency, along with education, training, and selectivity in assigning LEA staff handling AI tools. In addition to propositions of civic education, national and international collaborations and ensuring that AI capabilities by LEAs are up-to-date and at an arm's race with those of criminals. Interestingly, apart from some concerns about facial recognition, findings reveal concerns around how AI is being deployed, rather than the mere deployment of AI tools by LEAs. With this the study provides vital insights into the varied nature of measures that citizens deem necessary as safeguards to ensure their acceptance of AI.

**Acknowledgment**

This aspired work is based on the EU funded research project AIDA The AIDA project has received funding from the European Union's Horizon 2020 research and innovation program under grant agreement No 883596 (AIDA- Artificial Intelligence and advanced Data Analytics for Law Enforcement Agencies).